\documentclass[twocolumn, pra, aps, superscriptaddress,longbibliography,showpacs,amsmath,amssymb,floatfix]{revtex4-1}              
\usepackage{silence}
\usepackage{graphics}
\usepackage{subcaption}
\usepackage{amssymb}
\usepackage{amsmath}
\usepackage{epsfig}
\usepackage{color}
\usepackage{tabu}
\usepackage{booktabs}
\usepackage{mathtools}
\usepackage[colorlinks,linkcolor=blue,anchorcolor=blue,citecolor=blue,urlcolor=blue]{hyperref}
\usepackage{physics}
\usepackage{float}
\usepackage{diagbox}
\usepackage{ragged2e}
\usepackage{verbatim}

\begin{document}
\title{Interacting Mathieu equation, synchronization dynamics and collision-induced velocity exchange in trapped ions}
\author{Asma Benbouza}
\affiliation{CAS Key Laboratory of Quantum Information, University of Science and Technology of China, Hefei, 230026, People’s Republic of China}
\author{Xiaoshui Lin}
\affiliation{CAS Key Laboratory of Quantum Information, University of Science and Technology of China, Hefei, 230026, People’s Republic of China}
\author{Jin Ming Cui}
\affiliation{CAS Key Laboratory of Quantum Information, University of Science and Technology of China, Hefei, 230026, People’s Republic of China}
\affiliation{Synergetic Innovation Center of Quantum Information and Quantum Physics, University of Science and Technology of China, Hefei, Anhui 230026, China}
\affiliation{Hefei National Laboratory, University of Science and Technology of China, Hefei 230088, China}
\author{Ming Gong}
\email{gongm@ustc.edu.cn}
\affiliation{CAS Key Laboratory of Quantum Information, University of Science and Technology of China, Hefei, 230026, People’s Republic of China}
\affiliation{Synergetic Innovation Center of Quantum Information and Quantum Physics, University of Science and Technology of China, Hefei, Anhui 230026, China}
\affiliation{Hefei National Laboratory, University of Science and Technology of China, Hefei 230088, China}

\date{\today }

\begin{abstract}
Recently, large-scale trapped ion systems have been realized in experiments for quantum simulation and quantum computation. They are the simplest systems for dynamical stability and parametric resonance. In this model, the Mathieu equation plays the most fundamental role for us to understand the stability and instability of a single ion. In this work, we investigate the dynamics of trapped ions with the Coulomb interaction based on the Hamiltonian equation. We show that the many-body interaction will not influence the phase diagram for instability. Then, the dynamics of this model in the large damping limit will also be analytically calculated using few trapped ions. Furthermore, we find that in the presence of modulation, synchronization dynamics can be observed, showing an exchange of velocities between distant ions on the left side and on the right side of the trap. These dynamics resemble to that of the exchange of velocities in Newton's cradle for the collision of balls at the same time. These dynamics are independent of their initial conditions and the number of ions. As a unique feature of the interacting Mathieu equation, we hope this behavior, which leads to a quasi-periodic solution, can be measured in current experimental systems. Finally, we have also discussed the effect of anharmonic trapping potential, showing the desynchronization during the collision process. It is hopped that the dynamics in this many-body Mathieu equation with damping may find applications in quantum simulations. This model may also find interesting applications in dynamics systems as a pure mathematical problem, which may be beyond the results in the Floquet theorem. 
\end{abstract}

\maketitle 

\section{Introduction}
Mathieu's equation, which holds a prominent place in mathematical physics, has a  fascinating  historical origin dating back to the 19th century. Emile Mathieu conducted  research on the vibrations of elliptical drums. His work was motivated by the broader scientific context of understanding the wave phenomena, particularly in the study of acoustics and vibrations. In 1868, Mathieu published his seminal work and his investigations are rooted in the Helmholtz equation \cite{Ruby1996Applications,Bibby2013AccurateCO}
\begin{equation}
\nabla^2 W + k^2 W =0,
\end{equation}
in which the solutions are defined as $W(u,v,z) = f(u) g(v) \phi(z)$. In the elliptic cylinder coordinates $x = \rho \cosh(u) \cos(v)$, $y = \rho \sinh(u) \sin(v)$, and $z = z$. In the above equation, the curve by $u = \text{constant}$ gives the confocal ellipses and the curve by $v = \text{constant}$ gives the orthogonal hyperbolas. If we define $Y = f$ and $t = u$, or  $Y = g$ and $t = v$, then we will obtain the same ordinary differential equation, now known as the Mathieu equation, which emerged as a solution to describe the complex oscillatory patterns observed in elliptical hoops \cite{Mathieu1868Memoire}.
For this reason, the Mathieu and the modified Mathieu equation can emerge from any differential equations involving the Helmholtz equations expressed in the elliptic cylinder coordinates. It has been discussed in details by Paul in 1953 - 1958; see Ref. \cite{Paul1990Electromagnetic} and references therein. For a long time, the Mathieu equation has been one of the most important linear differential equations studied in mathematics and physics \cite{Landau1976Mechanics,Bhattacharjee2007AnIntroduction}, and can be written as the following second-order linear ordinary differential equation 
\cite{Richards1983TheMathieu,Paul1990Electromagnetic}
\begin{equation}
\frac{d^2Y}{dt^2} + (a - 2 q \cos(2t)) Y = 0.
\label{eq-mathieu}
\end{equation}
Here $t$ is the time, $Y$ is the vibration amplitude, and  the coefficients $a$ and $q$ known as the Mathieu parameters, completely determine the stability of the dynamics. Obviously, the differential operator itself is a periodic function, with period $T = \pi$, however the solution is not necessarily periodic, but instead, a quasi-periodic solution is allowed. This kind of quasi-periodicity is a typical feature of differential equations in dynamical systems \footnote{For example, let us consider the following ordinary differential equation $\left(\frac{d}{dx}+ A\sin(x)+B \right)y=0$, with solution $y = y_0 \exp(A \cos(x) - B x)$, which is not a periodic solution of $x$. However, it is a quasi-periodic solution satisfying $y(x + 2\pi) = \exp(-2\pi B) y(x)$, as expected from the Floquet theorem. }.

The above Mathieu equation is a special condition of the Hill equation, which can be written as 
\begin{equation}
\frac{d^2 Y}{dt^2} + (a +  2\sum_{j=1}^\infty q_j \cos(2j t)) Y =0,
\end{equation}
where $q_j$ are constants. The solution is related to a determinant of an infinite matrix. When only $q_1$ and $q_2$ are involved and $q_j =0$ for all $j >2$, it will be reduced to the Whittaker-Hill equation \cite{PARRAVERDE2024Steady,Urwin1970IIITheoryOf, Koenig1933Calculation}. 

In Ref. \cite{Ruby1996Applications}, a lot of important applications have been summarized, including elliptic drums, inverted pendulum, radio frequency quadrupole, modulated LC circuit, floating body {\it etc..}. In recent years, some major attention have been payed to the Paul trap for charged particles and mirror trap for neutral particles. In Refs. \cite{Daniel2020Exact,Alastair2014AnIntroduction,Cirac2000Ascalable,Steven2016Constructing}, applications of the Mathieu equation in wave propagation in pips, electromagnetic wave guides and oscillations of water in a lake have also been presented. In these applications, the physics in the Paul trap is the major concern in this work for its potential applications in quantum computation, following the scheme by Cirac and Zoller \cite{Cirac1995Quantum}. 

The Mathieu model provides a mathematical framework to understand the behavior of charged particles in various types of traps. Trapped ions are confined spatially by electric or magnetic fields, and their motion within these traps is described by the Mathieu equation \cite{Leibfried2003Quantum, Paul1990Electromagnetic}.  In this case, the dynamics are fully determined by several parameters: $a$, $q$ and the damping rate $\gamma$ (see below). Moreover, the trap geometry and field strength are also determined from the values of these parameters \cite{Savaryn2016Aresearchers, Paul1990Electromagnetic, Thompson2002Therotating}. Motivated by these researches, in this work, we extended the single particle Mathieu equation to the many-body Mathieu equation, and study its dynamics based on classical equations. This work is organized in the following way. The dynamics in the quantum regime need to be considered elsewhere. We first start by presenting the criterion for stability and instability based on Floquet theorem in Sec. \ref{sec-FloquetTheorem}. Then, in Sec. \ref{sec-stronggamma}, we study the dynamics with two to four trapped ions, in which in the large damping limit, the dynamics can be solved analytically. In this case, we show how the damping rate influences the relative phase and the vibration amplitude around their equilibrium positions. In Sec. \ref{sec-effectofdamping}, we study the dynamics with $N$ trapped ions with a finite damping rate. In Sec. \ref{sec-ManyBodyPhysics}, we examine the stability and the phase chart in the many-body Mathieu equation, which has been argued in the previous section. We show that the phase chart is independent of the many-body interaction. In Sec. \ref{sec-NONLINEAR}, the fate of the phase chart and the stability is examined in the presence of anharmonic nonlinear potential, showing that the parametric resonance is absent in the presence of a cubic nonlinear coefficient. Finally, we discuss the experimental relevance of our results in Sec. \ref{sec-Exp}. We conclude this work and discuss the unique feature of this many-body Mathieu model and its potential applications in Sec. \ref{sec-conc}. 

\section{Stability of the Mathieu equation and the phase chart}
\label{sec-FloquetTheorem}
The Mathieu equation is the simplest equation to study the stability and instability of a system with a periodic driving force, which is known as parametric resonance. This behavior is totally different from the forced resonance in the linear differential equation. In the forced resonance, the resonance can only happen when the driving frequency is the same as the intrinsic vibrational frequency of the system, in which the vibrational amplitude increases linearly with time $t$. However, in the parametric resonance, the resonance can be found in a wide range of parameters, with a vibrational amplitude increasing exponentially with time $t$ \cite{Landau1976Mechanics, Bhattacharjee2007AnIntroduction, Wu2012Mathieu}. In the following, we discuss the phase chart of the Mathieu equation with and without damping \cite{Trypogeorgos2016Cotrapping}. 

\begin{figure}[H]
    \centering \includegraphics[width=0.5\textwidth]{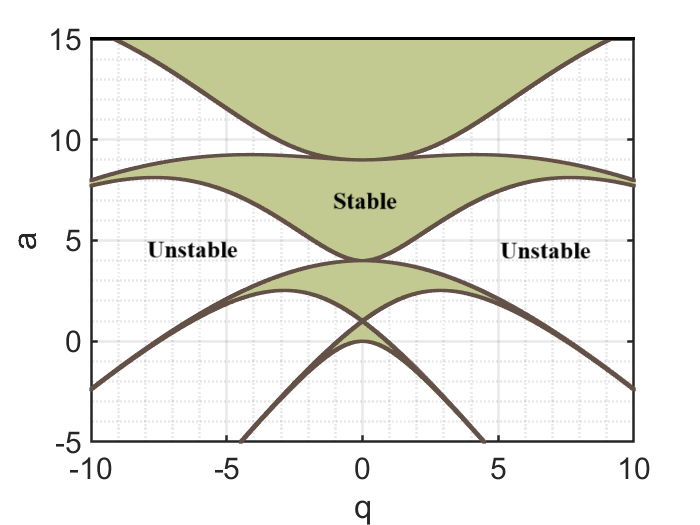}
    \caption{Stability chart with stable and unstable regions (tongues). The shadowed regimes correspond to the condition $|\beta|>1$ in Eq. \ref{eq-eqEigVal} without damping. }
    \label{fig-fig1}
\end{figure}

\begin{figure}[H]
\centering
\includegraphics[width=0.5\textwidth]{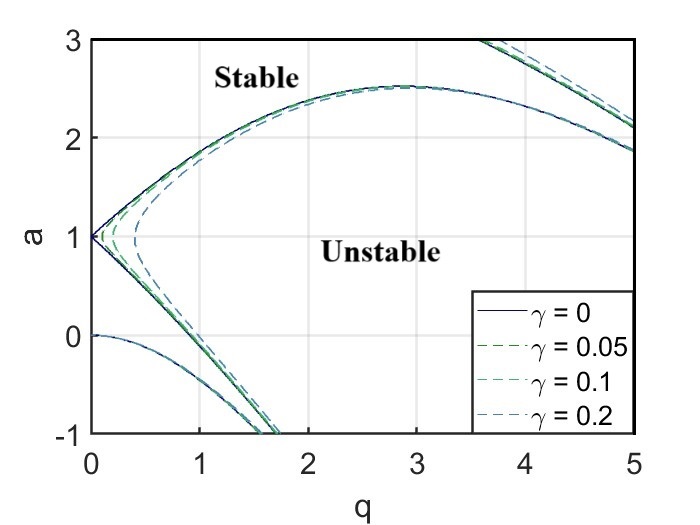}
\caption{Comparative analysis of stable and unstable regions of the damped and undamped Mathieu equations.}
\label{fig-fig2}
\end{figure}

We can use the Floquet theory to understand the stability of this model. Let us consider the following equation with periodic coefficients \cite{Kovacic2018Mathieu,Stafford1984Recent, Richards1976StabilityDiagramApprox}
\begin{equation}
\frac{d^2Y }{dt^2}+ 2\gamma \frac{dY}{dt} + (a - 2 q \cos(2t)) Y = 0, 
\label{eq-mathieut}
\end{equation}
where $\gamma$ is the damping rate \cite{Acar2016Floquet, Lawrence1993Thedamped}. This equation can be defined as  \cite{SINHA2001Parametric}
\begin{equation}
\begin{bmatrix}
\dot Y   \\
\ddot Y   \\
\end{bmatrix}
=
\begin{bmatrix}
0 & 1 \\
-  (a - 2 q \cos(2t)) & -  2 \gamma \\
\end{bmatrix}
\begin{bmatrix}
Y \\
\dot Y \\
\end{bmatrix}
= A 
\begin{bmatrix}
Y \\
\dot Y \\
\end{bmatrix}.
\end{equation}
We will obtain two independent solutions, that can be denoted as $(Y_{1}(t) , \dot{Y}_{1}(t))^T$, and $(Y_{2}(t) , \dot{Y}_{2}(t))^T$ , with the initial conditions
\begin{equation}
\begin{bmatrix} Y_{1}(0) \\ \dot{Y}_{1}(0) \end{bmatrix} = \begin{bmatrix} 1 \\ 0 \end{bmatrix}, \quad 
\begin{bmatrix} Y_{2}(0) \\ \dot{Y}_{2}(0) \end{bmatrix} = \begin{bmatrix} 0 \\ 1 
\end{bmatrix}.
\end{equation}
Furthermore, we can construct the solutions as 
\begin{equation}
\Psi=
\begin{bmatrix}
Y_{1}(t) & Y_{2}(t) \\
\dot{Y}_{1}(t) & \dot{Y}_{2}(t) \\
\end{bmatrix}, 
\end{equation}
with $\dot{\Psi}= A \Psi$, which based on the Floquet theory, its solution should be $\Psi(t + T) = \Psi(t) C$. One can check this solution $\dot{\Psi}(t+T) = A \Psi(t+T) = A \Psi(t) C = \dot{\Psi}(t)C$. Using the initial condition that $\Psi(0) = 1$, we can immediately obtain 
\begin{equation}
C = \begin{bmatrix}
Y_{1}(T) & Y_{2}(T) \\
\dot{Y}_{1}(T) & \dot{Y}_{2}(T) \\
\end{bmatrix}.
\end{equation}
In this way, we expect $\Psi(nT) =  \Psi(0) C^n$. Thus, the system is stable when and only when the eigenvalues of $C$, which are assumed by $\beta$,  satisfy $|\beta| \le 1$. We can obtain the eigenvalues of the $C$ matrix using 
\begin{equation}
    \beta^2 - \mathrm{Tr}(C) \beta +  \mathrm{det}(C) = 0.
    \label{eq-eqEigVal}
\end{equation}
The above equation is used to determine the phase boundary.  In the above solution, we have $Y_i \in \mathbb{R}$ and $\dot Y_i \in \mathbb{R}$,  thus $C$ is a real matrix; yet non-Hermite. (I) When the solution is real, we should have a phase boundary at $1 \pm \mathrm{Tr}(C) + \mathrm{det}(C)  = 0$; (II) When $\beta$ is a complex, the two solutions are complex valued, and we expect $\mathrm{det}(C) =1$.  The second condition will corresponds to the condition without damping $\gamma = 0$ \cite{Wilkinson2018Approximation, Jia2016OrderofParameters}. 

The phase boundary is fully determined by the two parameters $\mathrm{Tr}(C)$ and $\mathrm{det}(C)$.  We find that the phase boundaries are determined by the regime enclosed by the following condition
\begin{eqnarray}
1 \ge \text{det}(C)^2 \ge  |1 \pm \mathrm{Tr}(C)|^2.
\end{eqnarray}
With this theory, we can determine the phase boundary as a function of $a$ and $q$.  In Fig. \ref{fig-fig1}, we present the phase chart without damping \cite{Koad2022OnPreheating, Gavin2023novel}, and in Fig. \ref{fig-fig2}, we present the results with damping \cite{Fernando2018Lecture, Insperger2003Stability}. The roots of the equation without damping are given by
\begin{equation}
\beta = \frac{ \mathrm{Tr}(C) \pm \sqrt{ \mathrm{Tr}(C)^2 - 4  }}{2},
\end{equation}
and the stable phase is given by $|\mathrm{Tr}(C)| \le 2$ \cite{Taylor1969StabilityRegions,KONENKOV2002Matrix}. In this condition, $\beta$ is a complex number, and we can prove exactly that $|\beta | = 1$, in this condition we can define $\beta = \exp(i\theta)$. 

We see that the damping has the effect of reducing the unstable regions of the system, which is consistent with our intuition; see \cite{Kovacic2018Mathieu, Poulin2008stochastic} and Ref. \cite{Landau1976Mechanics}. Whenever there is a dissipation of energy, the system goes to rest, which naturally means that it becomes stable. However, in the following section, we will present a different role played by the dissipative term on the relation between the vibration amplitude and the equilibrium position in the strong damping limit.

\section{Dimensionless and large damping limits} 
\label{sec-stronggamma}

Now, we turn to the major results that will be presented in this work. Our aim is to study the dynamics of trapped ions in the presence of a strong Coulomb interaction. The Hamiltonian for $N$ trapped ions can be written as 
\begin{equation}
H = H_0 + U,
\end{equation}
where 
\begin{eqnarray} 
&& H_0 = \sum_{i=1}^N \frac{p_i^2}{2m} +\frac{1}{2} m \omega^2  ( a- 2q \cos( \Omega \tau )) q_i, \\ 
&& U =\frac{e^2}{4\pi \epsilon_0} \sum_{i < j} \frac{1}{|q_i -q_j|}.
\end{eqnarray}
In the above model, $m$ is the mass of the trapped ions, $\epsilon_0$ is the vacuum permittivity, $\Omega$ is the modulation frequency, and $\omega$ is the system's frequency.  The equation of motion of this model can then be written as 
\begin{eqnarray}
m\ddot{q}_i  = &&  - 2 \alpha  \dot{q}_i  - m \omega^2  (a-2q\cos(\Omega \tau))q_i \nonumber \\  
&& -\frac{2e^2}{4\pi \epsilon_0}  \sum_{j \ne i}\frac{1}{(q_i-q_j)^{2}}.
\end{eqnarray}
Here $\alpha$ is introduced for the damping effect. 

Now, we use the following transformation to make the equation  dimensionless
\begin{eqnarray} 
 \tau    &&  = \frac{2t}{\Omega}, \quad q_i =  \omega^2 \sqrt{\mathcal{Q}} y_i, \quad  \gamma   = \frac{\alpha}{m \omega}, \nonumber \\
  \omega &&  = \frac{\Omega}{2}, \quad \mathcal{Q} = \frac{2e^2}{4\pi\epsilon_0m \omega^3}.
\label{EqsUnitlessVar}
\end{eqnarray}
Substituting the new variables and parameters into the original equation results in the following interacting Mathieu equation
\begin{eqnarray}
   \ddot y_i && = -\hat{L}(y_i) - \sum_{j \ne i} \frac{1}{(y_i-y_j)^2}.
\label{eq:YwithF}
\end{eqnarray}
Hereafter, we have defined a differential operator as 
\begin{equation}
\hat{L}(y) = 2 \gamma \dot{y}  + (a-2q\cos(2 t))y.
\end{equation}
Now all the quantities became dimensionless, and their values are of the order of unity. In the last term, the minus sign represents the repulsive interaction between the charged ions. Obviously, when $2q > a$ and $\gamma$ is very small, the trapping potential may become unbounded from below, which may lead to instability.

\subsection{The case of two ions $N = 2$} 
The dynamics of a system of two ions can be effectively analyzed through the decomposition of the system into two components: the motion of the center of mass and the motion of the relative position of the ions. This decomposition simplifies the analysis by reducing the problem from a complex two-body interaction to two independent single-body equations. To this end, we define $2 R =y_1+y_2$, and $ r = y_1-y_2$. The system yields two equations
\begin{eqnarray}
    \ddot{R} \pm \frac{\ddot{r} }{2} + \hat{L}(R \pm \frac{r}{2}) - \frac{1}{r^2}=0. 
\end{eqnarray}
These two equations lead to two independent equations as following
\begin{eqnarray}
        && \ddot{R} + \hat{L}(R) =0, \quad \ddot{r} +\hat{L}(r) - \frac{2}{r^2}=0.
\end{eqnarray} 

\begin{figure}[H]
\centering
\includegraphics[width=0.45\textwidth]{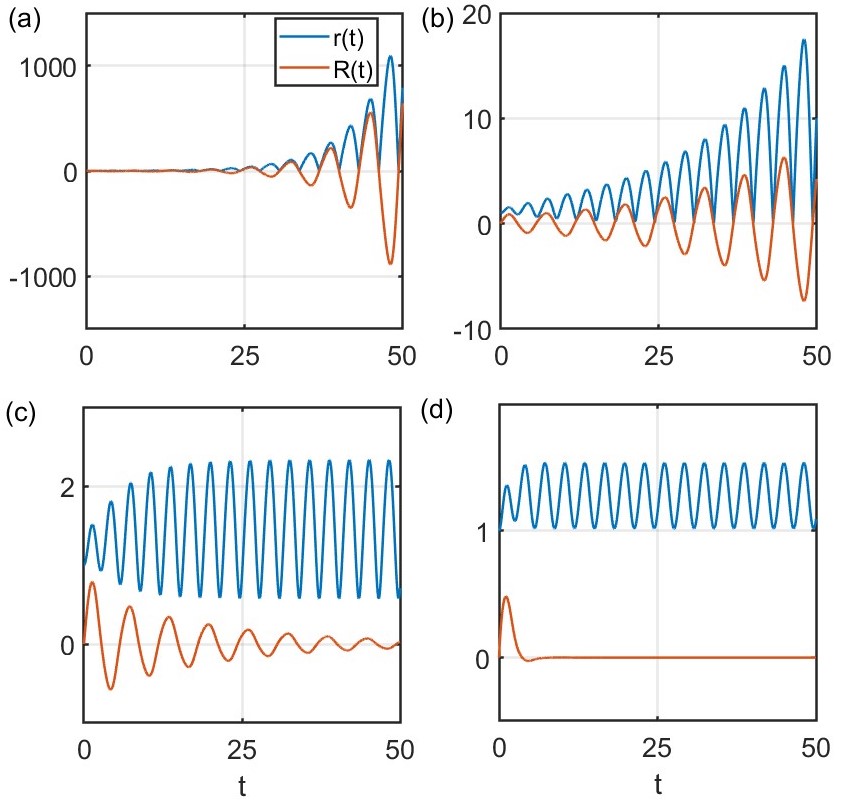}
\caption{\justifying Evolution of a system of 2 ions for a fixed pair of $a=1$, $q=0.3$. (a) - (d) represent the evolution with $\gamma = 0$, 0.1, 0.2 and 0.7, respectively. Since $r(0) = 1$, during the evolution, $r(t) > 0$ from the non-penetration effect of the charged ions. (a) and (b) are the dynamics in the unstable phase; and (c) and (d) are the results in the stable phase due to the damping. In the large $\gamma$ limit, the solution of $r$ can be described by Eq. \ref{eq-largegammar}. }
\label{fig-fig3}
\end{figure}

Here the first equation (the motion of the relative position) corresponds to the damped Mathieu equation, and the second one (the motion of the center of mass) corresponds to the same damped Mathieu equation but with a Coulomb interaction term. In the instability phase when $r \rightarrow \infty$, the center of mass will be reduced to the linear Mathieu equation. For this reason, these two equations have the same phase boundary for stability-instability transitions. However, in the stable phase, their dynamics will be totally different. We find that $R = 0$ is the solution of the center of mass, but $r = \text{constant}$ is not the solution of the relative position. 
For this reason, even in the presence of damping, the motion of the relative position will always oscillate in some way. Moreover, due to the non-penetration effect of the trapped ions, the motion of $r$ will be restricted to $r < 0$ or $r > 0$, depending on its initial value. The results for various damping rates are presented in Fig. \ref{fig-fig3}, showing that in the stable phase when $R \rightarrow 0$, $r$ will always oscillate periodically around its equilibrium position with the same period of the Mathieu equation. 

In the case of strong damping (see Fig. \ref{fig-fig3} (d)), we may assume $r = r_e + \delta r$, where $r_e$ is the equilibrium position, one may find that $\delta r$ oscillates in the way of the driving dynamics. When $\gamma \rightarrow \infty$, we expect $ar_e -2/r_e^2= 0$, yielding $r_e = (2/a)^{1/3} = (2)^{1/3}$ for $a =1$ 
used in Fig. \ref{fig-fig3}. Thus, we  assume that the dynamics are the following 
\begin{equation}
r = \left( \frac{2}{a} \right)^{1/3} + A \sin(2t + \theta) + \cdots,
\end{equation}
where $\theta$ is a phase introduced by the damping effect. 
Inserting this solution into the above equation, and to the leading term, we get
\begin{eqnarray}
 4\left(\frac{a}{2}\right)^{1/3} A\gamma - 2q\cos\left(\theta\right) &&=0, \\ 
\frac{3a  -4}{{\left(2/a\right)^{1/3}}}  A +2q \sin \left (\theta \right) &&= 0.
\end{eqnarray}
These two terms correspond to the coefficients of $\cos(2t)$ and $\sin(2t)$, which should be equal to zero. The higher-order terms such as $\cos(4t)$ and $\sin(4t)$ are neglected. The
solutions of $A$ and $\theta$ can be written as 
\begin{equation}
A = \frac{q}{2 \gamma (a/2)^{1/3}} \cos(\theta),
\end{equation}
and 
\begin{equation}
 \frac{ (3a-4)q}{2 \gamma} \cos(\theta) + 2q\sin(\theta)= 0.
\end{equation} 
In the large $\gamma$ limit, we expect $\theta \sim 0$, or
$\sin(\theta) = {(4-3a)}/{ 4\gamma}$. In this limit, we expect the solution to be 
\begin{equation}
r = \left( \frac{2}{a} \right)^{1/3} + \frac{q}{2^{2/3}a^{1/3}\gamma} \sin(2t).
\label{eq-largegammar}
\end{equation}

This solution can be understood by assuming 
$r =(2/a)^{1/3}+y$, where for small $y$ and $|(2/a)^{1/3}| \gg |y|$, we have
\begin{equation}
-\frac{2q\cos(2t)}{(a/2)^{1/3}} + (ay + \ddot y + 2\gamma \dot y -2q y \cos(2t)) = 0.
\end{equation}
For the four terms between the brackets, the term $2\gamma \dot y$ dominates and we have 
\begin{equation}
-\frac{2q\cos(2t)}{(a/2)^{1/3}} + 2\gamma \dot y = 0,
\end{equation}
yielding the following solution 
\begin{equation}
y = \frac{qr_e}{2\gamma} \sin(2t) = 
\frac{q}{2 (a/2)^{1/3}\gamma} \sin(2t),
\label{eq-qre}
\end{equation}
in Eq. \ref{eq-largegammar}. For this reason, the effect of damping influences the vibration amplitude, instead of driving the ions to rest. This picture applies to the conditions of more trapped ions. 

The above results can also be understood from the balance between the energy $E_\text{driving}$ introduced by the driving force and the energy $E_\text{damped}$ dissipated by the damping force. We have 
\begin{eqnarray} 
E_\text{driving} &&= \int_0^{\pi} -2q\cos(2t) x \dot{x} dt = -2 \pi A q r_e, \\ 
E_\text{damped} &&= \int_0^{\pi} 2\gamma  \dot{x}^2 dt = 4 \pi A^2 \gamma.
\end{eqnarray} 
A direct calculation using $E_\text{driving} + E_\text{damped} = 0$ will yield the above relation, with (see Eq. \ref{eq-qre})
\begin{equation}
A = \frac{q r_e}{2\gamma}.
\label{eq-Aqregamma}
\end{equation}
This method has been used in Ref. \cite{Landau1976Mechanics} in the determination of the vibration amplitude of the damped oscillator, and the same idea can be used to understand the vibration amplitude in our model. As a result, when $E_\text{driving} + E_\text{damped} > 0$ the vibration amplitude will increase due to the absorption of the energy from the driving field; otherwise, it will decrease. This picture can be used to understand the quasi-periodic motion of the trapped ions, as discussed below. Thus, the quasi-periodic solution can also be found in nonlinear driven equations. 

\subsection{The case of three ions $N = 3$}
The same method can be applied to three trapped ions, by defining $ r_1= y_1-R$, $r_2= y_3-R$, and $3R=y_1+y_2+y_3$. This results in a system of three coupled equations
\begin{eqnarray}
&& \ddot R  -  \ddot r_1 -\ddot r_2  + \hat{L}(R - r_1 - r_2)   \nonumber  \\
&&  -\frac{1}{(-r_1-2r_2)^2}+\frac{1}{(2r_1+r_2)^2}=0, \\
&&    \ddot R + \ddot r_1 + \hat{L}(R + r_1)  \nonumber \\
&& -\frac{1}{(r_1-r_2)^2}-\frac{1}{(2r_1+r_2)^2}=0,  \\
&& \ddot R + \ddot r_2+\hat{L}(R+r_2)  \nonumber \\
&& +\frac{1}{(-r_1-2r_2)^2}+\frac{1}{(r_1+r_2)^2}=0.
\end{eqnarray}
With this definition, we can obtain the equation of the center of mass as $\ddot{R} + \hat{L}(R) =0$, which is exactly the same as the driving Mathieu equation. This conclusion can be extended to arbitrary $N \ge 2$ ions. The equations of motion for $r_1$ and $r_2$ can be written as 

\begin{eqnarray}
    \ddot r_1 + \hat{L}(r_1) =&&  \frac{1}{(r_1-r_2)^2} +\frac{1}{(2 r_1 + r_2)^2} ,\\
     \ddot r_2 + \hat{L}(r_2) =&& - \frac{1}{(r_1-r_2)^2} -\frac{1}{( r_1 + 2 r_2)^2}.
\end{eqnarray}

\begin{figure}[H]
\centering
\includegraphics[width=0.45\textwidth]{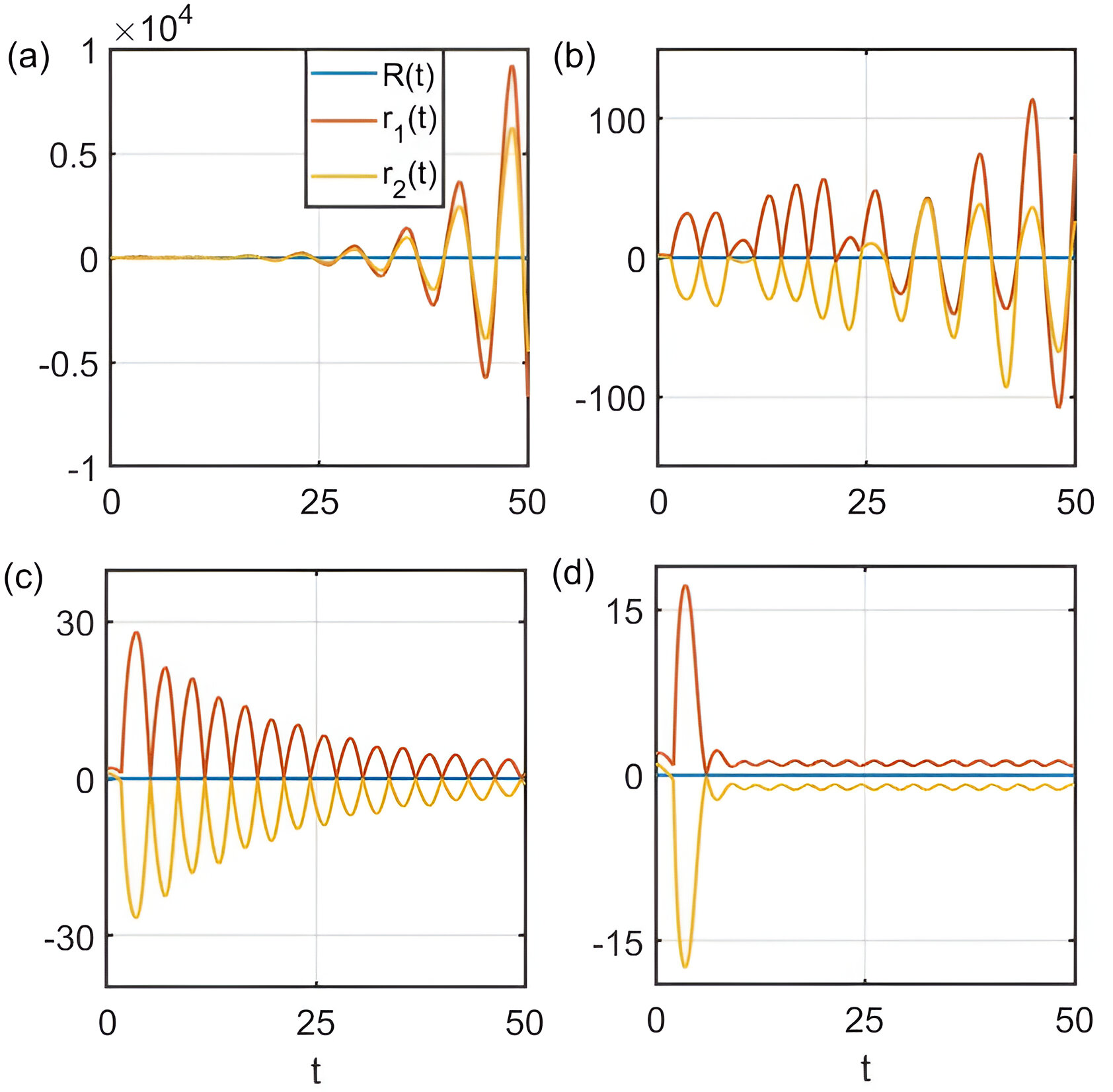}
\caption{\justifying Evolution of a system of 3 ions for a fixed pair of $a=1$, $q=0.3$. (a) - (d) represent the evolution with $\gamma = 0$, 0.1, 0.2 and 0.7, respectively. }
\label{fig-fig4}
\end{figure}

We first start by defining the equilibrium position equations for a strong damping $r_1= r_e + \delta r$, and $ r_2= -r_e - \delta r$. In the large $\gamma$ limit, we neglect $\delta r$, and would obtain  
\begin{equation}
- a r_e = {1\over 4r_e^2} + {1\over r_e^2} \rightarrow r_e= \frac{5^{1/3}}{2^{2/3}a^{1/3}}.
\label{eq-reN3}
\end{equation}
Therefore, the solutions would be
\begin{eqnarray}
    r_1=-r_2= \frac{5^{1/3}}{2^{2/3}a^{1/3}} + A\sin(2t+\theta).
    \label{eq-sin2ttheta}
\end{eqnarray}
Using the perturbation theory, and assuming $A$ to be small, we obtain from its equation of motion that
\begin{eqnarray}
4 A\gamma - 2q\frac{5^{1/3}}{2^{2/3}a^{1/3}} \cos(\theta) &&=0, \\ 
\frac{(5a -4 ) q \cos(\theta)}{2 \gamma} + 2 q \sin(\theta)&&= 0.
\end{eqnarray}

The solutions of $A$ and $\sin(\theta)$ can be written as 
\begin{eqnarray}
 A =  \frac{5^{1/3} q \cos(\theta)}{ 2 a^{1/3} 2^{2/3} \gamma}, \quad \sin(\theta)= \frac{(4-5a)}{4 \gamma} \cos(\theta).
\end{eqnarray}
In the large $\gamma$ limit, $\theta \sim 0$, expecting the solution to be 
\begin{equation}
r = \frac{5^{1/3}}{2^{2/3}a^{1/3}} + \frac{5^{1/3} q }{ 2 a^{1/3} 2^{2/3} \gamma} \sin(2t).
\label{eq-largegammar3}
\end{equation}
Obviously, the vibration amplitude in this solution also satisfies the energy balance condition of Eq. \ref{eq-Aqregamma}. This solution has been verified using numerical data. In Fig. \ref{fig-fig4}, we present the results for three trapped ions, showing that in the large $\gamma$ limit, indeed, when $R \rightarrow 0$, $r_1$ and $r_2$ will vibrate around their equilibrium positions, which can be well described by the above analytical solution 
\begin{eqnarray}
 2 \gamma \dot y -\frac{ 5^{1/3} 2 q }{2^{2/3} a^{1/3} } \cos(2t) =0,
\end{eqnarray}
which yields $y=\frac{5^{1/3} q }{2^{2/3}2 a^{1/3} \gamma} \sin(2t)$; see Eq. \ref{eq-largegammar3}. 

\subsection{The case of four ions $N =4$}
This method can be extended to any arbitrary number of trapped ions, and here we will only present results for $N=4$. In a similar way, for $i={1,4}$ and $j={2,3}$ we can define 
\begin{eqnarray} 
 y_i = \pm r_e^1 \pm A_1 \sin(2t), \quad  y_j = \pm r_e^2 \pm A_2 \sin(2t).
\label{eq-y1to4}
\end{eqnarray}
This result is based on numerical simulation, showing that in the large $\gamma$ limit, the trapped ions 1 and 4 (and 2 and 3) can have almost opposite dynamics. From the equation of motion and assuming $A_i$ to be small in comparison to $r_e^i$, we obtain
\begin{equation}
r_e^1 = {1.4368 \over a^{1/3}}, \quad 
r_e^2 = {0.454379 \over a^{1/3}},
\end{equation}
which  can not be solved analytically. We also find using numerical fitting that 
\begin{equation}
A_1 = 1.4368 \frac{q}{2\gamma a^{1/3}}, \quad 
A_2 = 0.454379 \frac{q}{2\gamma a^{1/3}},
\end{equation}
which can be understood using Eq. \ref{eq-Aqregamma}. From these results, we expect that in the condition of $N \ge 4$, we should have $r_e^i \propto 1/a^{1/3}$ and $A_i \propto q/a^{1/3}\gamma$, up to some universal constants to be determined numerically. 

So far, we focused on studying systems in the strong damping limit, and what we noticed is that it have barely any influence on the total energy of the ions. This means that despite the presence of damping from the environment or the laser field, the total energy of the system remains relatively conserved. The damping does not extract a significant large energy from the system, that is why it remains stable over time; see discussion in the next section. As a result, in this regime, the strong damping only influences the oscillation amplitudes of the ions around their equilibrium positions. In the following, we will only focus on the dynamics with finite damping, which will yield some new interesting features.

\subsection{Anharmonic effect} 
\begin{figure}[H]
\centering
\includegraphics[width=0.45\textwidth]{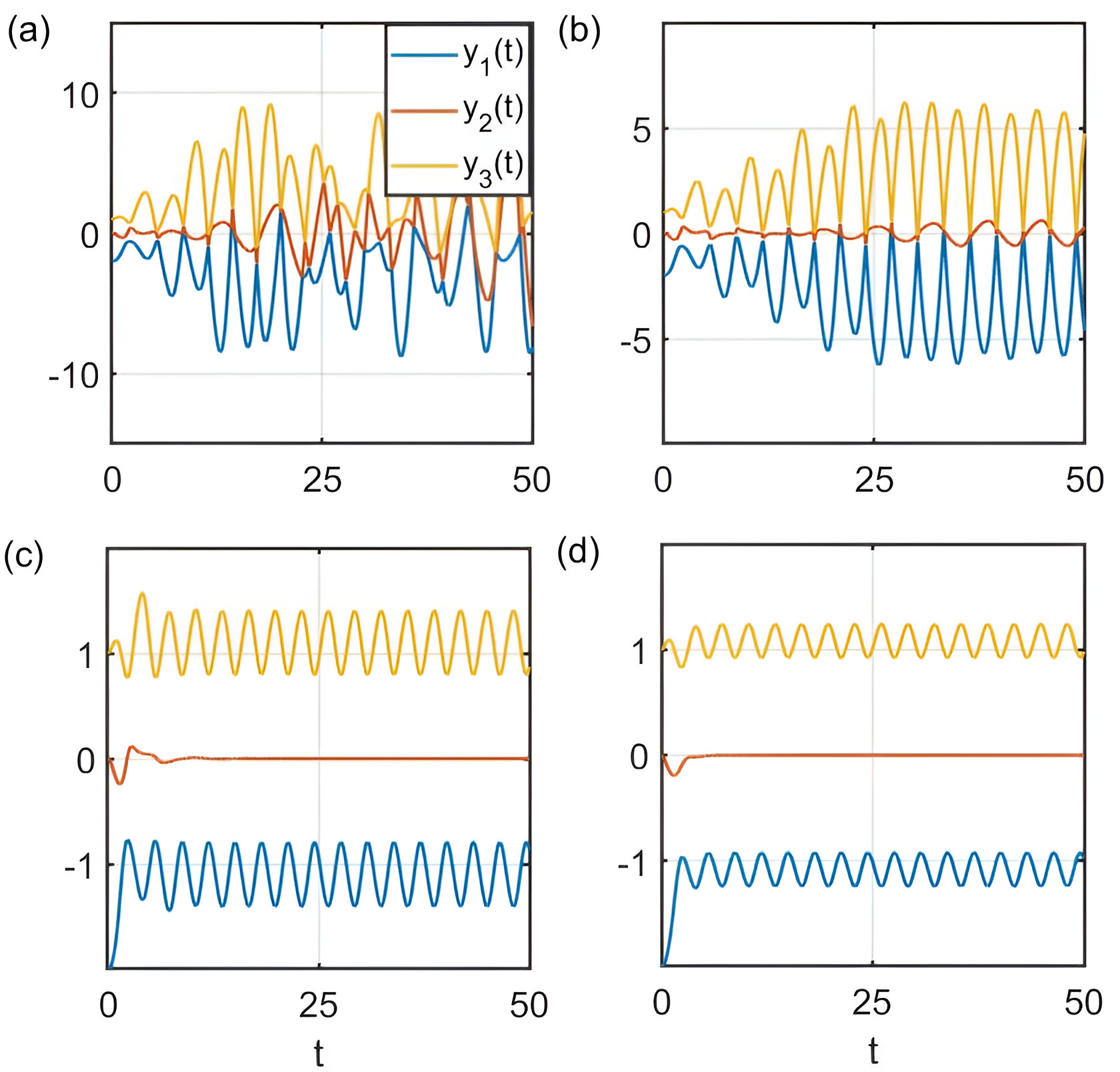}
\caption{\justifying Evolution of a system of 3 ions for  $a=1$, $q=0.3$, $\eta=0.01$. (a) - (d) represent the evolution with $\gamma = 0$, 0.1, 0.5 and 1, respectively. }
\label{fig-fig5}
\end{figure}

\begin{figure*}[!t]
\centering
\includegraphics[width=0.93\textwidth]{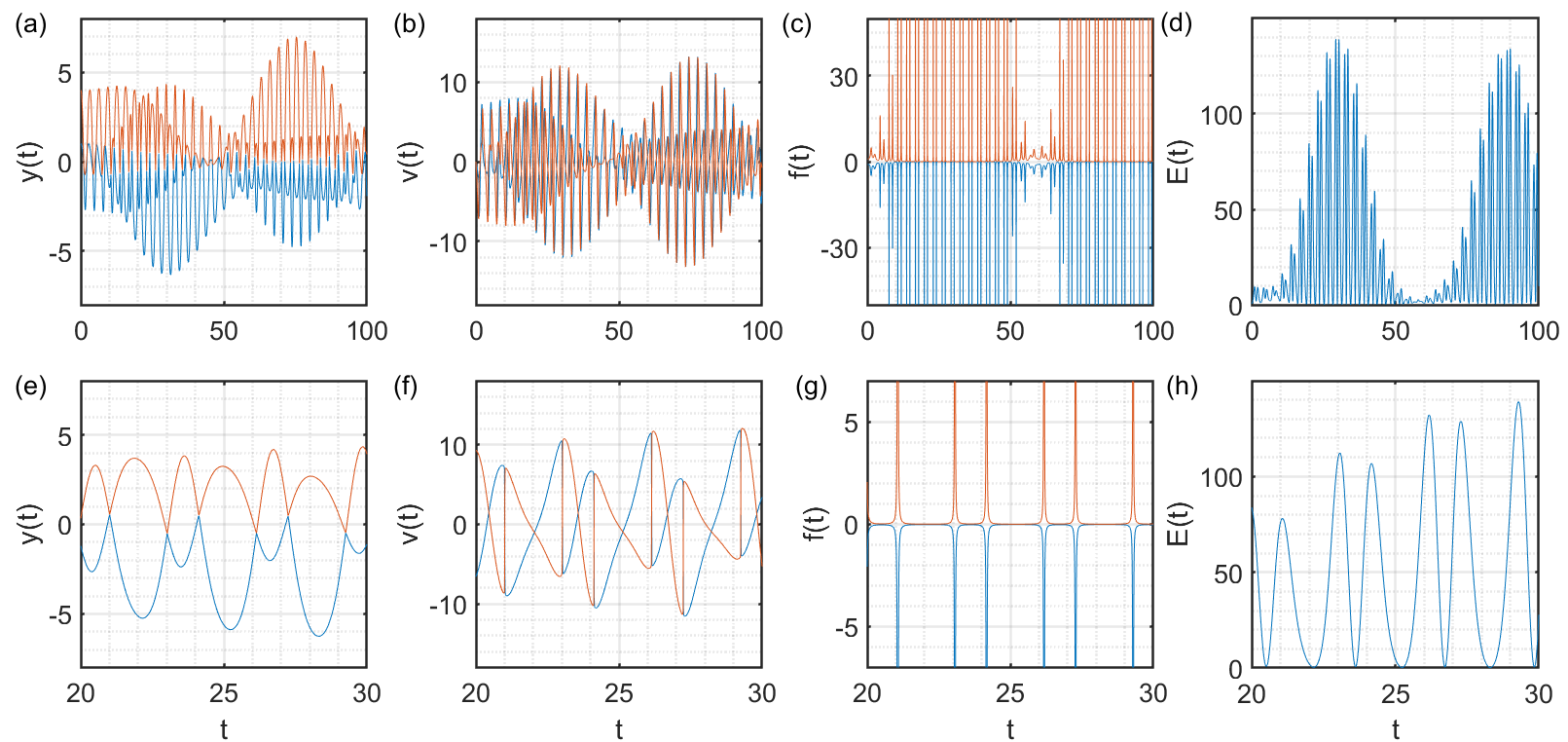}
\caption{\justifying The dynamics evolution of a system of 2 ions with asymmetric initial conditions, with $a=5$, $q=1.7$ and without damping. (a) and (b) depict a detailed view of the ions’ positions and velocities over a duration of $t=100$. (c) and (d) represent the evolution of the force and the energy over a period of time of $t=100$. Meanwhile, (e), (f), (g) and (h) depict the positions, the velocities, the forces and the energy of the ions over a shorter period of time $t=[20-30]$, respectively. In (g), the force diverges when the distance between the ions approach zero simultaneously from the synchronization dynamics. }
\label{fig-fig6}
\end{figure*}

The above analysis --- based on the center of mass and the relative coordinates --- depends strongly on the linearity of the equation of motion. In the presence of anharmonic effects in the trapping potential, the calculation is much more complicated. It has some profound impact on the dynamics of the system. Firstly, the method of the center of mass and the relative positions can not be applied anymore; and secondly, the phase chart and stability can be changed even in the nonlinear Mathieu equation; and thirdly, the criterion for stability based on the Floquet theorem will no longer be valid. For these reasons, the quasi-periodic solution may not be observed. 

To this end, let us consider the following model with $N = 3$ with strong damping. The equation of motion can be written as 
\begin{equation}
\ddot{y}_i  + \hat{L}(y_i) + 
\eta y_i^3 - \sum_{j \ne i} \frac{1}{(y_i-y_j)^2} = 0, \quad i = 1 -3. 
\end{equation}
Let us assume that $ y_{1,3} = \pm r_e \pm A_{1,3} \sin(2t)$, and $y_2 = 0$, where by symmetry the equilibrium position of the first and third ions should be at $\pm r_e$ and the equilibrium position of the second ion should be at zero. This solution is supported by numerical evidences in Fig. \ref{fig-fig5}. In the case of strong damping $\gamma$, $|A_{1,3}| \ll |r_e|$, thus we have 
\begin{equation}
ar_e + \eta r_e^3 = {5 \over 4 r_e^2}.
\label{eq-areeta}
\end{equation}
This equation is the same as Eq. \ref{eq-reN3} when $\eta = 0$, and will be solved using numerical methods. Using perturbation theory again, we will find the leading terms of the equations (with $A_1=-A_3$)
\begin{eqnarray}
4 \gamma A_{1}  - 2 q r_e = 0,
\end{eqnarray}
for the coefficient of $\cos(2t)$, allowing us to obtain the expressions of
\begin{eqnarray}
     A_1= -\frac{q r_e}{2 \gamma }, \quad 
     A_3= \frac{ q r_e}{2 \gamma }.
\end{eqnarray}
The coefficient of $\sin(2t)$ is neglected. A much better approximation of the solutions is using Eq. \ref{eq-sin2ttheta}, with a finite relative phase. These solutions are the same as Eq. \ref{eq-Aqregamma}. The equilibrium position $r_e$ can be numerically determined using Eq. \ref{eq-areeta}. Obviously, with the increasing of $\eta$ (assuming $\eta > 0$), $r_e$ will decrease monotonically. In the large $\eta$ limit, we should have 
$r_e \sim (5/4\eta)^{1/5} \simeq 1/\eta^{1/5}$. 

We see that in the condition of anharmonic effects in the trapping potential, we still have the relation in Eq. \ref{eq-Aqregamma} for the balance between the driving force and the damping force averaged in one full period. Therefore, the anharmonic effects, $\eta$, in the trapping potential only influences the vibration amplitude in a way of Eq. \ref{eq-areeta}. A simple and straightforward argument will show that this conclusion to be correct even with many trapped ions.  

\section{Exchange of velocities and Newton's cradle} 
\label{sec-effectofdamping}
\subsection{Without damping}

\begin{figure}
\centering
\includegraphics[width=0.5\textwidth]{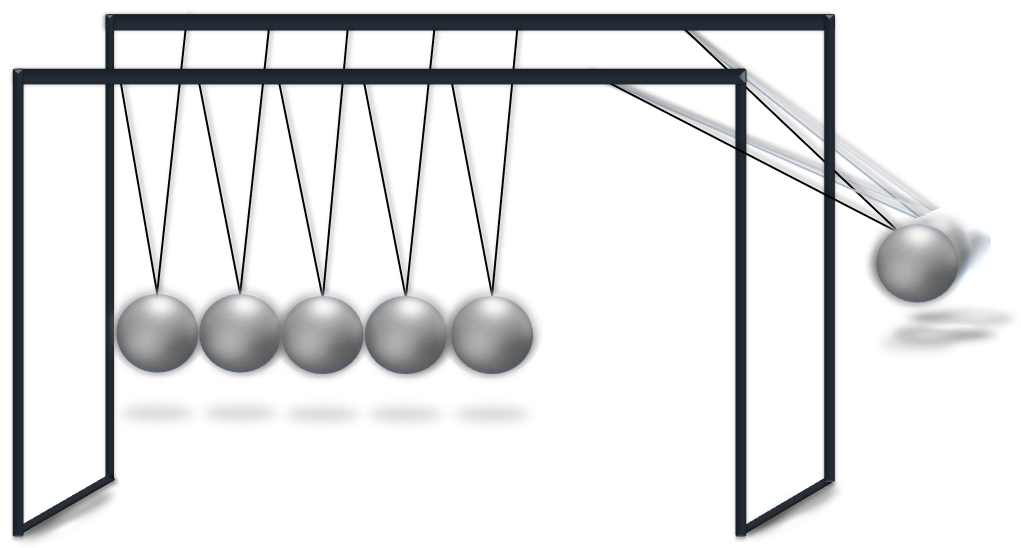}
\caption{\justifying Newton's cradle and the sudden exchange of velocities  between distant balls during collisions when all of them have the same mass and geometry. }
\label{fig-fig7}
\end{figure}

Then, what will happen in the presence of finite damping?  Generally speaking, the dynamics and vibration amplitudes depend on the competition of the driving forces, which may introduce some energy to the system, and the damping force, which can extract energy from the system. In the large $\gamma$ limit, these two forces are balanced, leading to periodic vibrations. In the long time limit, we find that the dynamics of the trapped ions are independent of their initial positions and velocities (see Eq. \ref{eq-y1to4}). In the weak damping limit, the physics will be totally changed. For example, in the model without damping, this system should support a quasi-periodic solution, instead of a periodic solution, and large vibrations of the amplitude of the ions will be possible, in which the distance between the ions may become very small, leading to large forces from the repulsive Coulomb interaction. In this way, the dynamics will exhibit some new features, which resembles to the dynamics of Newton's cradle; see Fig. \ref{fig-fig7}. This motivate our following investigations. 

We will focus on the following quantities: the positions 
$ y_i$, the velocities $v_i = \dot{y}_i$,  the forces $f_i = -\nabla_i U$, and the energy $E = T + U$. Here, $E$ is the total energy of the system  defined by the kinetic energy, the trapping energy and the Coulomb energy. When two ions approach each other, the force may diverge, leading to dramatic changes of velocities --- exchange of velocities  --- \cite{Van2022rfinduced}, which is the same as the dynamics in Newton's cradle. Furthermore, when the total energy $E$ increases, the system absorbs energy from the driving force; otherwise, the driving force will extract energy from the system.  

\begin{figure*}[!t]
\centering
\includegraphics[width=0.93\textwidth]{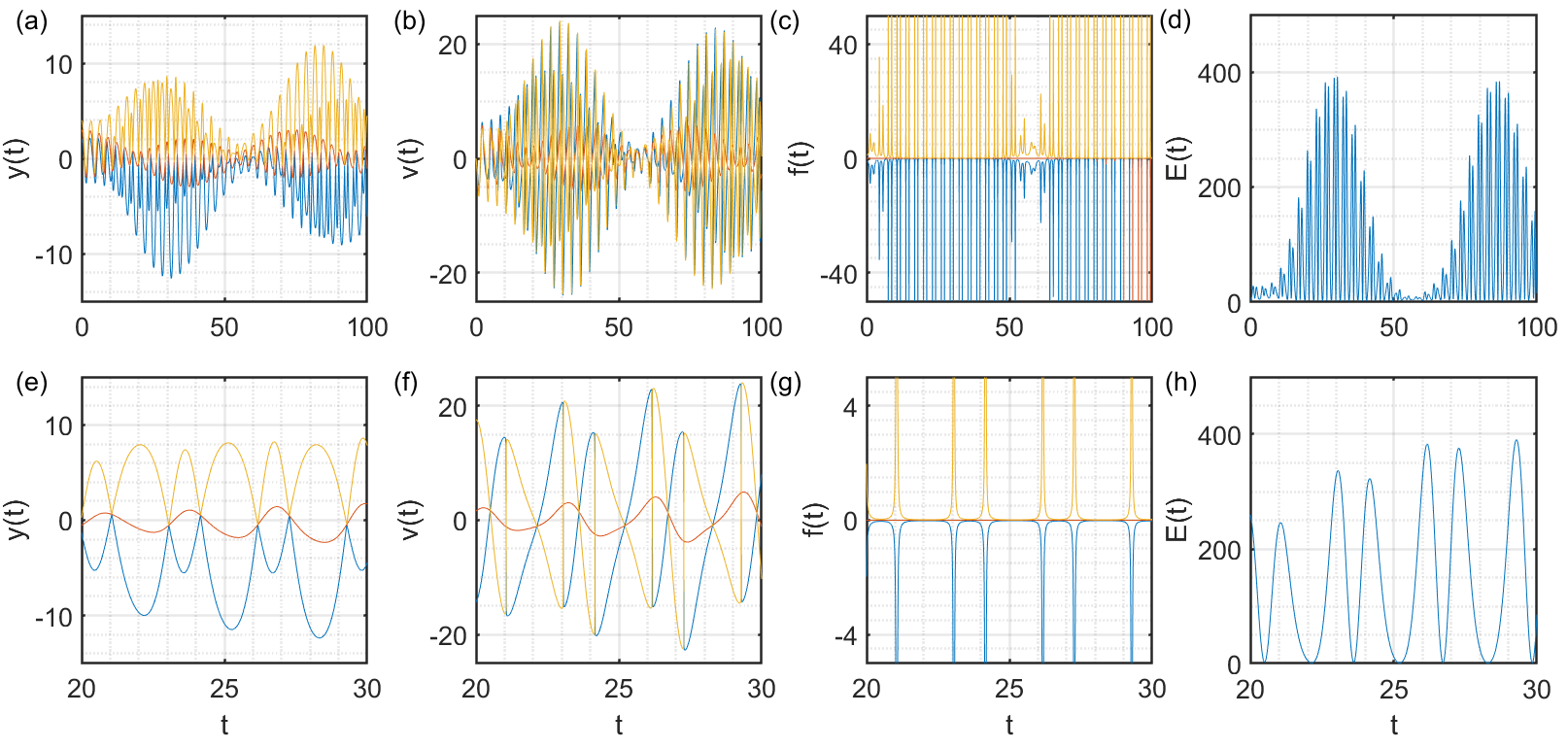}
\caption{\justifying The dynamics of a system of 3 ions with asymmetric initial conditions with $a=5$, $q=1.7$ and without damping. The meanings of the sub-figures are the same as that in Fig. \ref{fig-fig6}. }
\label{fig-fig8}
\end{figure*}

\begin{figure*}[!t]
\centering
\includegraphics[width=0.93\textwidth]{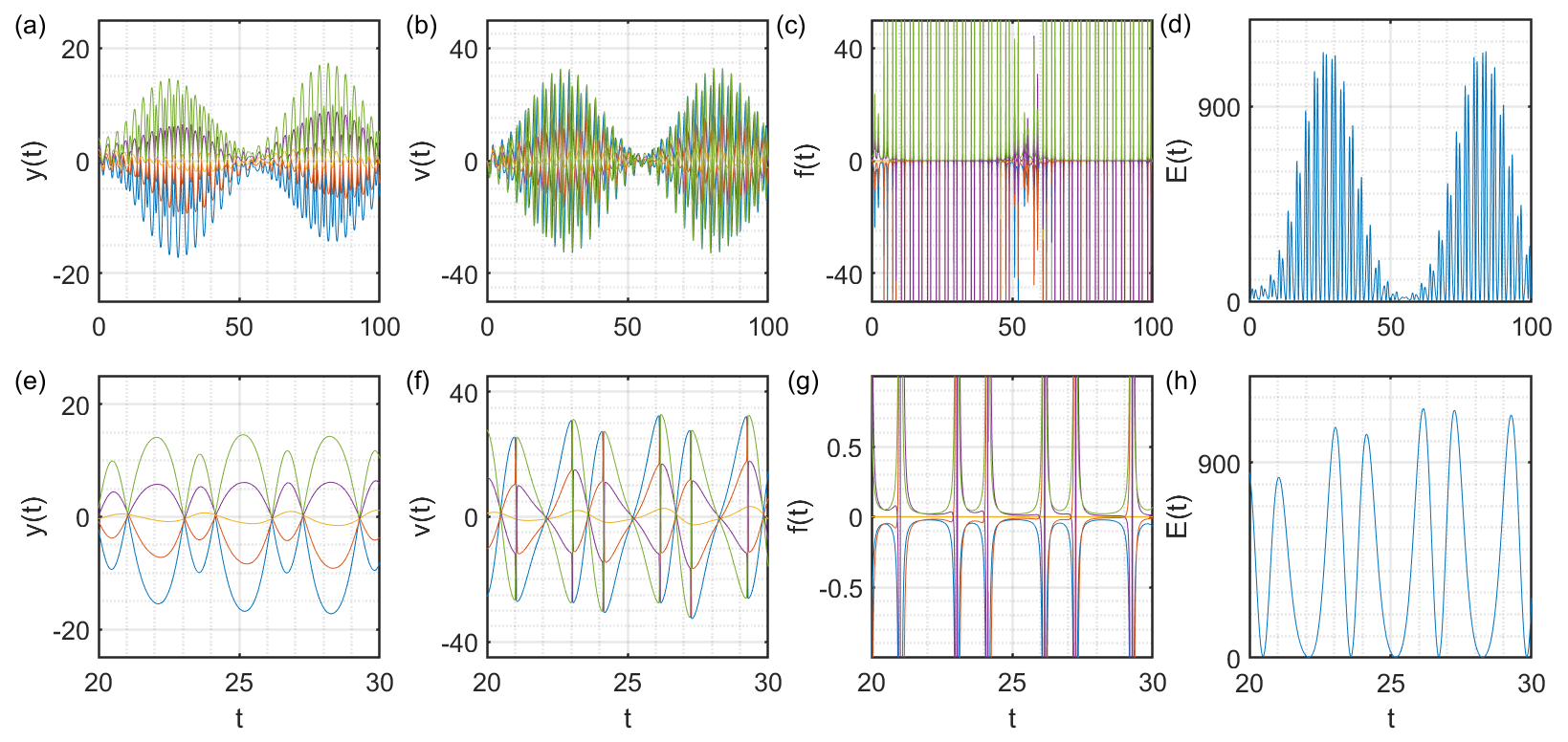}
\caption{\justifying The dynamics of a system of 5 ions with asymmetric initial conditions with $a=5$, $q=1.7$ and without damping. The meanings of the sub-figures are the same as that in Fig. \ref{fig-fig6}. }
\label{fig-fig9}
\end{figure*}

The results for $y_i$, $v_i$, $f_i$ and $E$ for two trapped ions are presented in Fig. \ref{fig-fig6} without damping. We focus on the stable regime. The dynamics of $y_1$ and $y_2$ exhibit some interesting features not presented in the strong damping limit.  Firstly, it presents some kind of micromotion with period $\pi$ from the driving field; and secondly, the system shows some global structure with much larger period, not necessary to the commensurate with the period of the driving field. In this way, the system displays some quasi-periodic dynamics. The velocity also exhibits similar dynamics in Fig. \ref{fig-fig6} (b). We find that the vibration amplitude and the velocity are greatly enhanced, indicating that the system continuously absorbing and releasing energy from the driving field. In Fig. \ref{fig-fig6} (f), we 
plot the detailed dynamics of the velocity, showing that when two ions are approaching each other (see Fig. \ref{fig-fig6} (e)), there is a sudden exchange of velocities, which is the same as that in Newton's cradle (Fig. \ref{fig-fig7}). Our model may be the smallest system for the realization of Newton's cradle. At this point, the force will become divergent or significantly large; see Fig. \ref{fig-fig6} (c) and (g). In Fig. \ref{fig-fig6} (d), we calculate the total energy of this model, showing that the total energy has the same quasi-periodic structure for the motion and the velocity. These results represent some unique features of the interacting Mathieu equation with the many-body Coulomb force. 

\begin{figure*}[!t]
\centering
\includegraphics[width=0.93\textwidth]{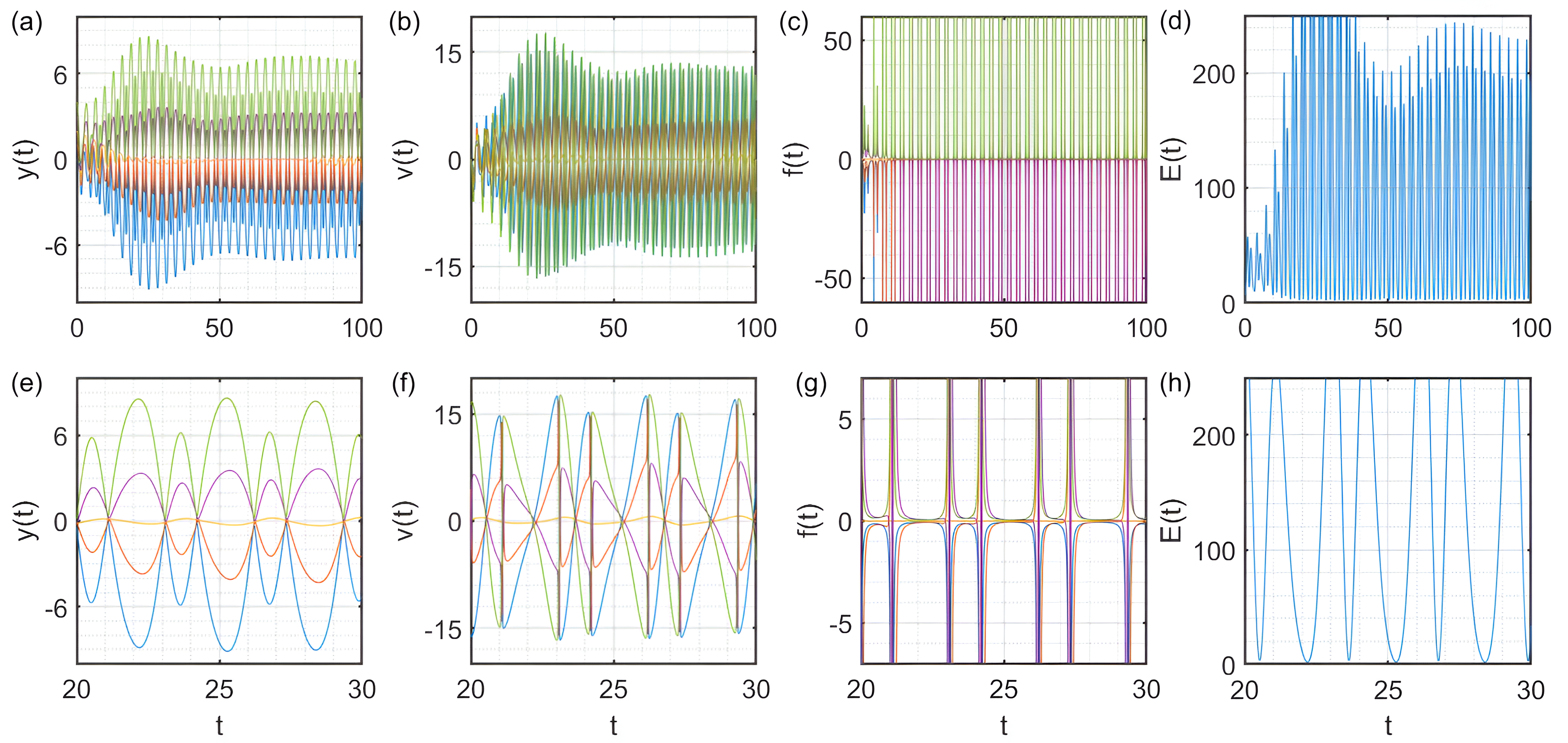}
\caption{ \justifying The dynamics of a system of 5 ions with asymmetric initial conditions with $a=5$, $q=1.7$, and weak damping  $\gamma =0.05$. The meanings of the sub-figures are the same as that in Fig. \ref{fig-fig6}. }
\label{fig-fig10}
\end{figure*}

Next, we are in a position to understand what will happen in the presence of many trapped ions. In Newton's cradle, the exchange of velocities  happens between two colliding balls; and the more number of balls involved the more complicated the dynamics become. To this end, we present the results for three trapped ions in Fig. \ref{fig-fig8} and for five trapped ions in Fig. \ref{fig-fig9} without damping. In these two figures, we always find the sudden exchange of velocities  when all the ions are colliding with each other. The most striking feature is that in the presence of many trapped ions ($N \ge 3$), the motion of the ions will be synchronized in such a way that they will collide to each other almost at the same time. We have verified that similar features can be found for much larger systems (for example, $N \ge 10$), and these synchronization dynamics are independent of their initial positions and initial velocities. As a result, for three trapped ions, the exchange of velocities  happens between $y_1$ and $y_3$ (assuming $y_1 < y_2 < y_3$); and for five trapped ions, the exchange of velocities  happen between $y_1$ and $y_5$, and $y_2$ and $y_4$. The synchronization dynamics are a unique feature in our model. 

These results should not be mixed up with the melting of ions, in which the structure is totally destroyed \cite{Van2022rfinduced}. In this case, the distance between the ions will increase, leading to the destruction of the crystal structures. In our model, we indeed found the dramatic increase of the distance between the ions, however, these ions are still trapped in the potential, and their distances will be decreased after some proper time, leading to coherent quasi-periodic dynamics. In this way, the structure of the ions will not be destroyed. 

\subsection{Effect of damping}
\label{subsec-damping}
With the above dynamics, we need to understand the role of the damping in the many-body Mathieu equation, in which we expect a transition from the undamped physics discussed in the above subsection to the strong damping physics discussed in the large $\gamma$ limit. These physics are intriguing not only because of the effect of damping on the exchange of velocities, but also from the transition from a quasi-periodic solution to a periodic vibrational solution. 

\begin{figure*}[!t]
%\centering
\includegraphics[width=0.93\textwidth]{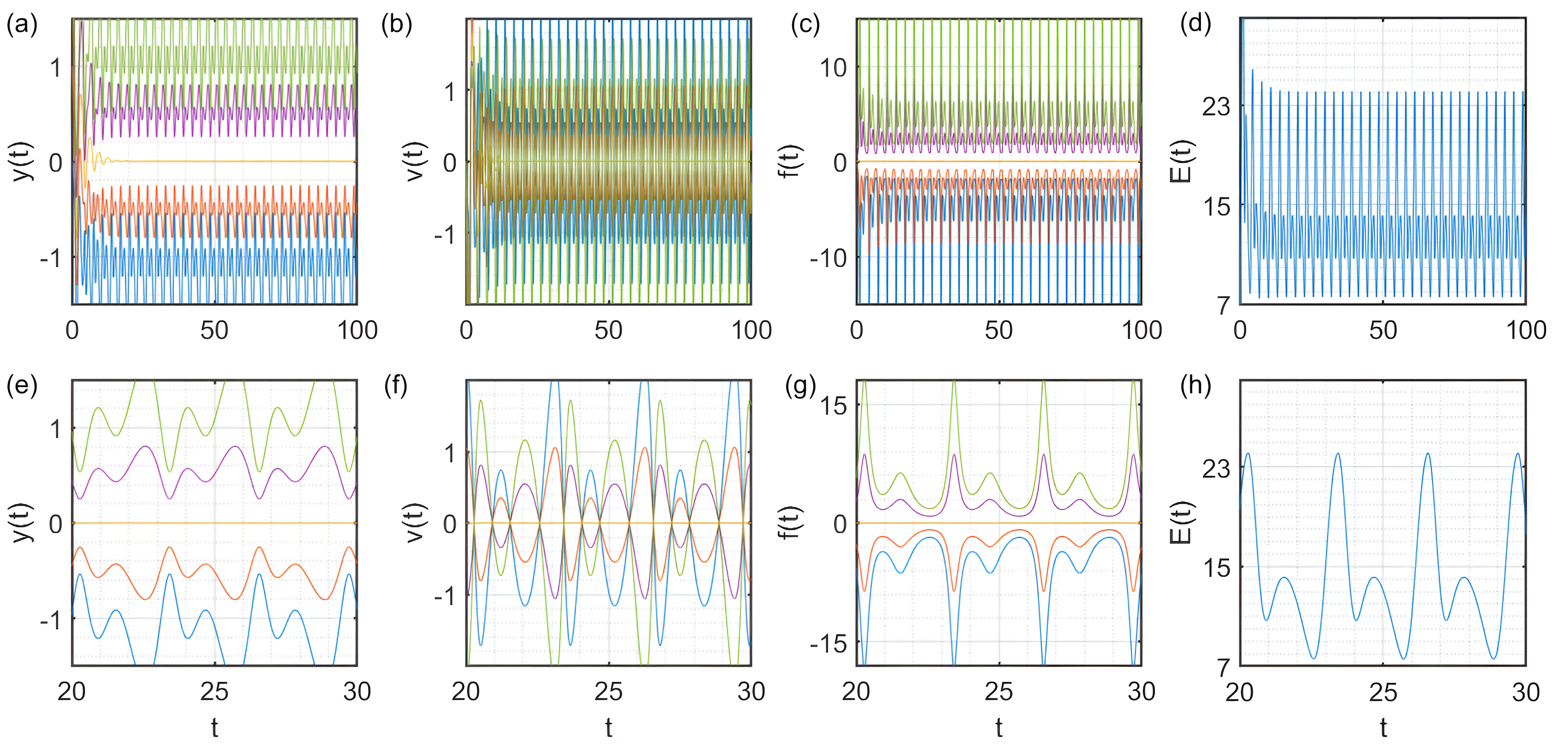}
\caption{\justifying The dynamics evolution of a system of 5 ions with asymmetric initial conditions with $a=5$, $q=1.7$, and a strong damping $\gamma =0.4$. The meanings of the sub-figures are the same as that in Fig. \ref{fig-fig6}. }
\label{fig-fig11}
\end{figure*}

By focusing on the dynamics of a system of five ions, we can observe much more the ion-ion interactions and highlight the unique features of damping. In the stable regime, the results for $y_i$, $v_i$, $f_i$ and $E$ for five trapped ions with a weak damping and a strong one are presented in Fig. \ref{fig-fig10} and Fig. \ref{fig-fig11}, respectively. In the case of weak damping, the system behaves almost exactly the same as without damping. This is seen in Fig. \ref{fig-fig10} (a) and (e), the ions oscillate in a quasi-periodic way where they approach each other, the Coulomb force becomes strong in Fig. \ref{fig-fig10} (c) and (g), leading to a sudden exchange of velocities , as observed before. In this condition, each ion will mirror the furthest ion in the system, thus we observe that ion $y_1$ exchanges its velocity with ion $y_5$, and $y_2$ exchanges its velocity with ion $y_4$, leaving ion $y_3$ to be almost unchanged. This is a unique feature from Newton's cradle, in which the kinetic energy is transferred from the left furthest ball to the right furthest ball through collisions, and everything in between stays stable. This feature is not changed significantly by the weak damping. 

However, with the increasing of the damping rate, say $\gamma=0.4$, as shown in Fig. \ref{fig-fig11}, we will notice some total different dynamics. We find that the previous structure of the motion of ions is destroyed; see Fig. \ref{fig-fig11} (a) and (b). Ions will no longer oscillate quasi-periodically, because the damping will introduce some stability to the system, and each ion now tends to oscillate around its equilibrium position. This will cause the  amplitude of each ion to be small in the large $\gamma$ limit. As a result, we will no longer have any exchange of velocities, as demonstrated in Fig. \ref{fig-fig11} (b) and (c). Now the dynamics of the system become more stable, and in this case, the distance between the trapped ions is large enough for the Coulomb force to be insignificant. The exchange of velocities  will be observed only when the distance between the ions is approaching zero. In the strong $\gamma$ limit, their dynamics will be reduced to that discussed in Sec. \ref{sec-stronggamma}, showing that the anharmonic effect term $\eta$ only influences the equilibrium position $r_e$, with relation between vibrational amplitude and equilibrium position given by Eq. \ref{eq-Aqregamma}. In this work, the validity of this relation is found in many different conditions. 

\section{Many-body physics stability chart}
\label{sec-ManyBodyPhysics}
\begin{figure}
\centering
\includegraphics[width=0.49\textwidth]{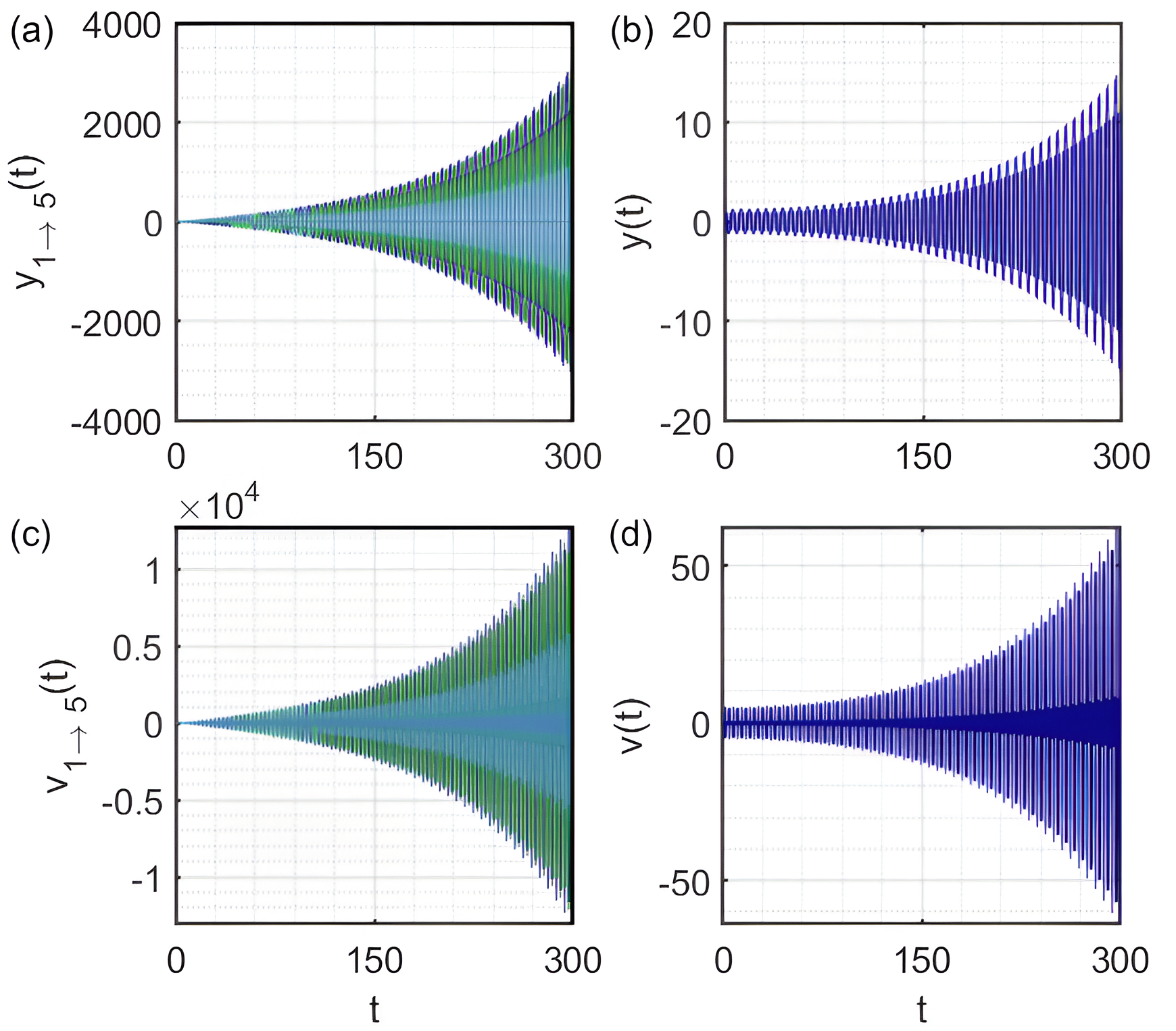}
\caption{\justifying Evolution of the system over a long period of time for a fixed pair of $a=15$, $q= 9.148$ in the unstable region, without damping ($\gamma = 0$). (a) and (c) are the results for 5 ions, and (b) and (d) are the results for 1 ion, respectively. }
\label{fig-fig12}
\end{figure}

\begin{figure}
\centering
\includegraphics[width=0.49\textwidth]{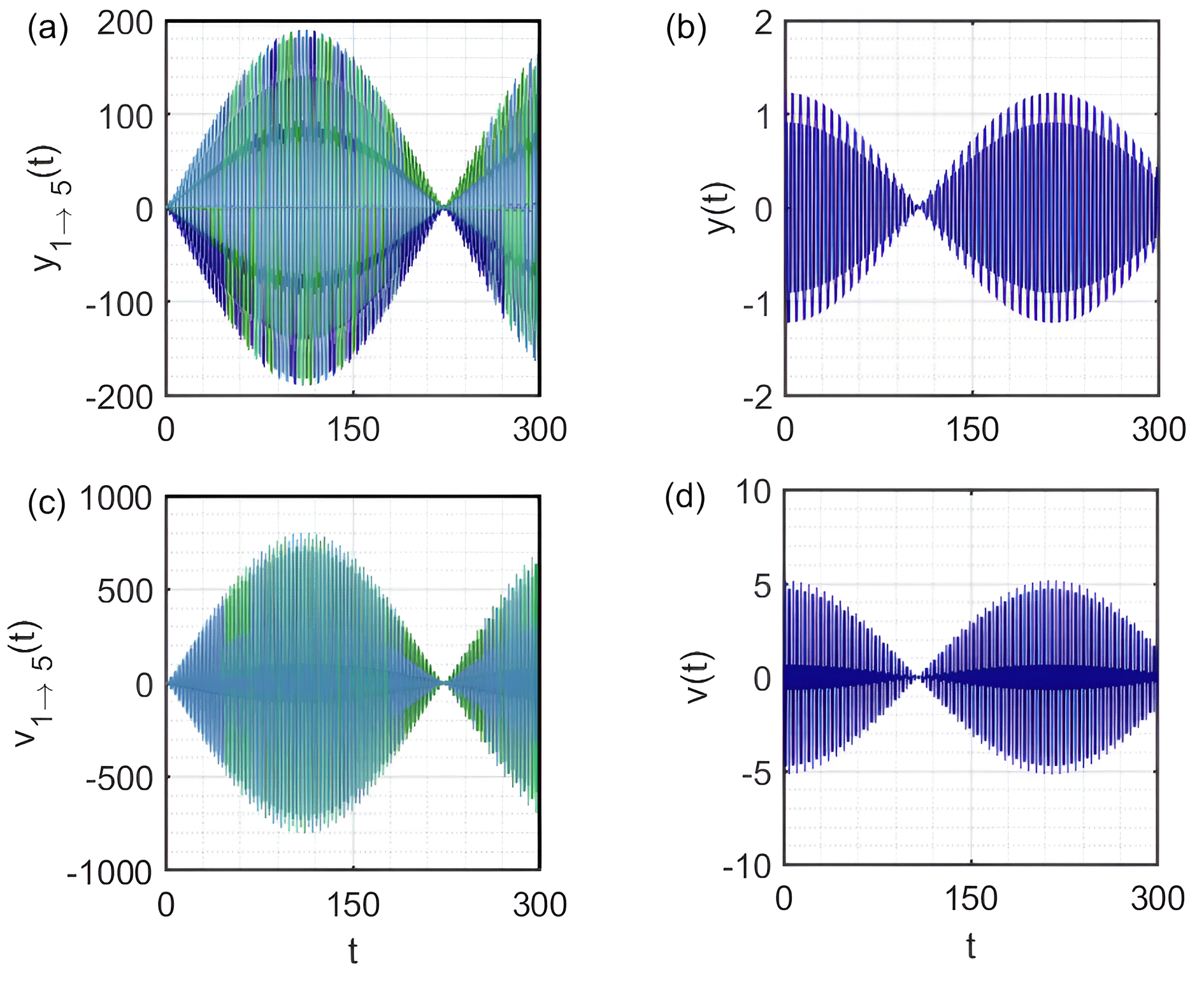}
\caption{\justifying Evolution of the system over a long period of time for a fixed pair of $a=15$, $q=9.146$ in the stable region without damping ($\gamma = 0$). The sub-figures depict the same information as Fig. \ref{fig-fig12}.}
\label{fig-fig13}
\end{figure}

Finally, we are in a position to understand the role of Coulomb interaction on the stability and the phase chart in this model using the following argument. Let us assume that the system is
unstable. In this case, $y_i$ approaches infinity, as observed in the single particle Mathieu equation. When the distance between the ions is sufficiently large, the Coulomb interaction is negligible, thus the instability of this model should be fully determined by the phase chart of the single particle Mathieu model. Thus, we have sufficient reasons to believe that the phase chart in this interacting Mathieu model should be the same as the single particle Mathieu equation. This is shown in Fig. \ref{fig-fig12}, however, the dynamics can still be fundamentally different. In Fig. \ref{fig-fig12}, we focus on the motion and the velocity in the unstable region for two groups of ions. The first group is composed of 5 ions in Fig. \ref{fig-fig12} (a) and (c), and the second group is composed of single ion in Fig. \ref{fig-fig12} (b) and (d). Both systems will oscillate exponentially, but have different magnitudes. The difference between the amplitude of the motion of the first ion and the fifth ion is around $10^{2}$, however, the difference between their velocities is around the order of $10^{3}$. The vibration 
magnitude of the position and the velocity are much smaller in a single particle model. Despite of this difference, both systems display similar patterns of oscillations. When a group of ions are sufficiently far from each other, the Coulomb forces decrease rapidly to zero, resulting in no influence on their motion. In such cases, the ions act like independent particles, which gives rise to a collective behavior similar to that of a single particle. 

Next, we study the motion and the velocity of 5 ions and of 1 ion in the stable region, which are presented in Fig. \ref{fig-fig13}, with $a = 15$ and $q_c \simeq 9.1473$. Similar features have also been observed. We find that these two conditions have the same parameters and they have the same stability. However, in the many-body model, the vibration amplitude is greatly enhanced by about two orders of magnitude, as compared with the single particle one. This result also means that in the many-body model the larger $N$ is, the larger the vibration amplitude will be. For this reason, the trapping of $N$ particles (where $N$ is large) without damping will become a significant issue. However, in the presence of damping, the enhanced vibration amplitude will be greatly suppressed. 

This observed phenomenon is fascinating in physics, because of the fact that the collective interactions of multiple particles give rise to behaviors similar to those of individual particles. This parallelism highlights not only  the fundamental principles regarding the behavior of trapped ion systems, but also the  interconnections of their dynamics. We hope these physics can be observed in experiments \cite{Van2022rfinduced}. In the case of strong damping, the single particle will cease to oscillate, yet the many-body one will oscillate around its equilibrium position, with position and amplitude satisfying Eq. \ref{eq-Aqregamma}. 

\section{Mathieu equation with cubic anharmonic term}
\label{sec-NONLINEAR}

The Mathieu equation with a cubic nonlinear term, in the presence of many trapped ions can be written as \cite{ELDIB2024AnInnovative, Pagano2018longionchain}
\begin{equation}
\ddot{y}_i  + \hat{L}(y_i) + 
\eta y_i^3 - \sum_{j \ne i} \frac{1}{(y_i-y_j)^2} = 0, 
\label{eq-NonLinear}
\end{equation}
where $\eta$ is the anharmonic coefficient. The cubic nonlinear coefficient introduces a nonlinearity to the equation. Therefore, the response of the system will no longer be directly proportional to the applied force, which naturally will make the system exhibit much more complicated behaviors. The cubic nonlinear coefficient in the Mathieu equation for trapped ions typically originates from the anharmonicity in the trapping potential that is responsible of the interaction between the ion's motion and the trapping field \cite{Galal2020Stability, Kidachi1997Note}. Numerical calculations show that in the presence of a weak $\eta$, the vibration amplitude is
\begin{equation}
A \propto {1\over \sqrt{\eta}},
\end{equation}
thus for all $a$ and $q$ the system is always stable. This relation can be determined by the minimal of $a/2 A^2 + \eta/4 A^4$, with $a < 0$, yielding $A^2 = -a/\eta$. Hence the cubic nonlinear term can fundamentally change the fate of the phase chart discussed before. In the following, we are interested in the effect of this nonlinear term on the vibration amplitude with strong damping. 
\begin{figure}
\centering
\includegraphics[width=0.49\textwidth]{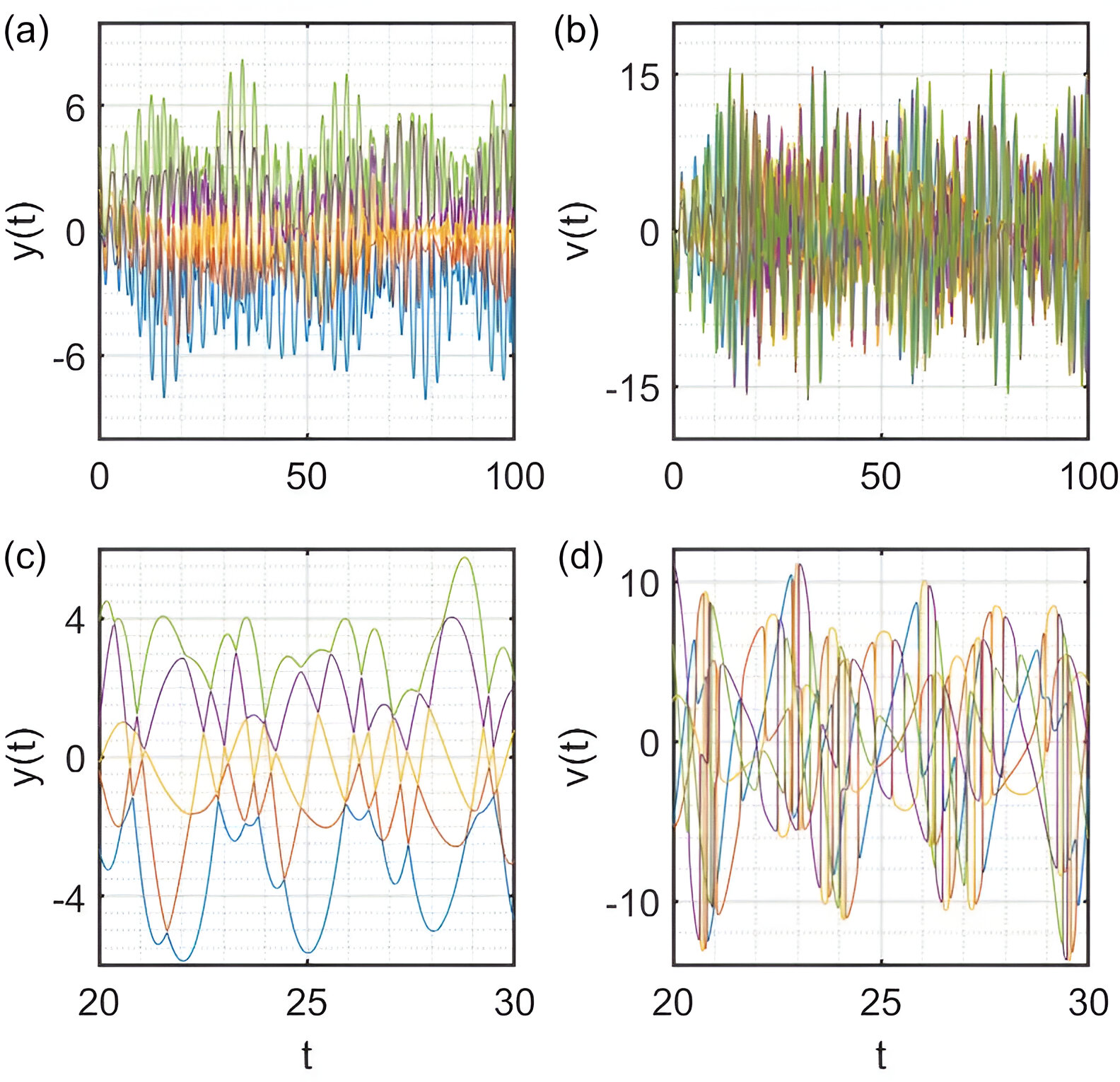}
\caption{\justifying Dynamics evolution of a system of 5 ions with asymmetric initial conditions in the stable region, with $a=5$, $q=1.7$, $\gamma = 0.0$ and  $\eta=0.01$. }
\label{fig-fig14}
\end{figure}
In this section, we will explore the effect of the nonlinear term on the collisions of ions. Assume $A$ to be the vibration amplitude, then we require 
\begin{equation} 
\eta A^2 \sim a \sim \mathcal{O}(1),
\end{equation}
in which condition the nonlinear term is important. By solving Eq. \ref{eq-NonLinear}, with $\eta=0.01$, we present the dynamics of 5 trapped ions in Fig. \ref{fig-fig14}. While the pattern of the ions looks organized, the pattern of the velocities is not. As compared with Fig. \ref{fig-fig9}, we find that the quasi-periodic dynamics no longer exists. Moreover, the exchange of velocities  between the first and the last ion will not be observed; instead, the collision of ions will happen at different times (see Fig. \ref{fig-fig14} (c) and (d)), leading to an exchange of velocities between adjacent ions. For this reason, the synchronization dynamics are not observed anymore. This is a typical feature in a nonlinear trap \cite{Pagano2018longionchain}, in which the ions in different positions may feel some slightly different trapping frequency, therefore their dynamics and period will be different. We find that when $\eta \sim 0.001$, the synchronization and the associated exchange of velocities  can still be observed. However, in the much stronger nonlinear trap, the dynamics will become much more complicated. Furthermore, in the case of strong damping, the ions will vibrate around their equilibrium positions, leading to periodic dynamics, as observed in Sec. \ref{sec-stronggamma}. In this case, the positions of ions may appear to have different vibration amplitudes and different effective periods. Thus, the synchronization comes from the parametric driving field and the harmonic potential, in which all ions tend to have the same behavior, including the same relative phase, which maybe measured in the future experiments. 

\section{Experiments and possible signatures} 
\label{sec-Exp}

When working with numerical simulations, it is necessary to use dimensionless variables to simplify the equations and reducing the complexity of the problem. However, to interpret these numerical results in real life, it is crucial to convert these dimensionless variables back into dimensional terms, and estimate their values. Therefore, we will have to estimate the values of $a$, $q$ and $\gamma$ possibly to be used in experiments. 

In a Paul trap, there is typically both a dc and an RF voltages applied to the electrode. Therefore, the total voltage  should be $V(t) = V_\text{dc}  + V_\text{RF} \cos(\Omega t)$. Created by this total voltage is the electric potential, which in one dimension reads as \cite{Trimby2022Reaching, Noshad2009Computation, Romaszko2020Engineering, Zhao2008Molecular,Poindron2023Thermal}
\begin{eqnarray}
    \Phi(y,t) = \frac{V_\text{dc} }{2 R_\text{dc}^2} y^2 + \frac{V_\text{RF}}{2 R_\text{dc}^2} y^2 \cos(\Omega t),
\end{eqnarray}
then, the electric field can be defined as $\textbf{E}= -\grad \Phi$, which will allow us to find the equation of motion of the ions using $\textbf{F}= e \textbf{E}$, yielding \cite{Ziaeian2010Theoretical, CHU1998Observation, Drewsen2000Harmonic, Paul1990Electromagnetic, NOSHAD2011Numerical, ZIAEIAN2011Theoreticalstudy}.
\begin{eqnarray}
m \ddot{y} = - 2  \alpha  \dot{y} - \frac{e }{R_\text{dc}^2}(V_\text{dc}  + V_\text{RF} \cos(\Omega \tau))y.
\label{EqVoltages}
\end{eqnarray}
Therefore, the expression of the parameters in the dimensional system \cite{Leibfried2003Quantum, Dawson2010Quadrupoles, Zhao2002Parametric, March2006Quadrupole, seddighi2012Study, Drewsen2000Harmonic, Dawson2010Quadrupoles}  
\begin{eqnarray}
    a && =\frac{4 e V_\text{dc} }{m \Omega^2 R_\text{dc}^2},  \quad  \abs{q}  =\frac{2 e V_\text{RF}}{m \Omega^2 R_\text{dc}^2 },\nonumber \\
    \gamma && = \frac{\alpha}{m \omega},    \quad t       = \frac{\Omega \tau }{2},  \quad  \Omega  = 2 \omega.
\end{eqnarray}
Here, $e$ is the charge of the ions, $V_\text{dc} $ is the applied dc voltage, $V_\text{RF}$ is the applied radio-frequency voltage,  $m$ is the mass of the ions, $\omega$ is the system frequency, $\Omega$ is the driving force frequency and $R_\text{dc}$ is a parameter that defines the geometry of the trap and represents the distance from the trap center to the electrode.

In an experimental setup for the trapped ions, the system is bound to experience some heating effect due to many reason, like collisions between the ions, exchange of energy with the environment, and fluctuations of the applied electric fields. Therefore, the trapped ions need to be cooled down, in order to maintain its stability. Many techniques can be used for this. In Refs. \cite{Van2022rfinduced, Ziaeian2012Theoretical, Kaplan09Laser}, two different mechanisms for the damping force were discussed, the first one is the drag force due to residual rarefied gas, and the other one is the interaction with the blue-detuned laser beams. In both cases, the damping rate $\gamma$ can be tuned in a wide range and in general the damping from the laser field dominates by Doppler effect. Especially, this damping force is essential for the cooling of trapped ions. That is why a damping term  was  introduced in Eq. \ref{EqVoltages}. Furthermore, $a$ and $q$ can be estimated knowing the values of experimental variables $V_\text{dc} $, $V_\text{RF}$, $e$, $m$ and $R_\text{dc}$. Using the above relations, we get
\begin{equation} 
\frac{a}{|q|}  =\frac{V_\text{dc} }{V_\text{RF}} ,
\label{eq-aoverq}
\end{equation}
which can be tuned in a wide range of experiments. Furthermore, using Eq. \ref{eq-Aqregamma}, the strong damping regime can be achieved when $|A| \ll r_e$, yielding 
\begin{equation}
|\frac{q}{2\gamma}| \ll 1.
\label{eq-qovergamma}
\end{equation}
Thus, we expect the physics discussed in this work to be researched using trapped ions. For example, in experiment with trapped $^{171}$Yb$^+$ ions with $\Omega = 
2\pi \times 5$ kHz \cite{Chen2022Neural, Trimby2022Reaching}, with $V_\text{dc} =100$ V, $R_\text{dc} = 5$ mm, $hc = 1.24$ eV$\cdot$$\mu$m, and $mc^2=171 \times 1860 \times 0.511$ MeV, and $\Omega = 2\pi \times 0.25$ MHz, we can obtain 
\begin{equation}
a = 3.5886.
\end{equation} 
The similar magnitude of $q$ can be obtained using Eq. \ref{eq-aoverq} by changing of $V_\text{RF}$. Much more details can be found in Ref. \cite{Trimby2022Reaching}. 
Obviously, these values can be tuned in a wide range in experiments, thus the phase chart can be studied using this platform. From Eq. \ref{eq-qovergamma}, we require $|\gamma| \gg 1$, assuming $|q| \sim \mathcal{O}(1)$, and from $\alpha \sim \omega$, we expect that the cooling of the trapped ions can be reached in the time scale of $\mathcal{O}(1/\omega)$ in the strong damping limit; see Fig. \ref{fig-fig5} and Fig. \ref{fig-fig11}. 

\section{Conclusions}
\label{sec-conc}

To summarize, we present a complete analysis of the interacting Mathieu equation with the Coulomb interaction, which can be realized using trapped ions. We investigate the physics in various conditions. In the strong damping limit, the dynamics can be solved using perturbation theory, showing that the damping coefficient can play the role of reducing the vibration amplitude about their equilibrium positions. In the weak damping limit, we observe the synchronization dynamics and the associated exchange of velocities  between distant trapped ions, which resembles to the dynamics of Newton's cradle. In this case, the phase chart for the stability and instability dynamics is independent of the Coulomb interaction, thus, it is the same as the single particle Mathieu equation. Finally, we study the effect of the nonlinear anharmonic term in the trapping potential, in which the system will always be stable for all the parameters. In this case, the desynchronization prohibits the exchange of velocities between distant ions. We estimated that these results are within the reach of the current experiments with trapped ions and expect these interesting dynamics to be observed. 

During the preparing of this work, we were constantly asking the fundamental question of what are the most unique features in this many-body classical model? The generalization of the exact solvable single particle Mathieu equation to the many-body models and even to the quantum realms are quite obviously important and necessary \cite{Leibfried2003Quantum, Singer2010trappedions}, which should exhibit some interesting physics. In most of cases, the quasi-periodic solutions are the major concern in these investigations. From these investigations with the Coulomb interaction in this work, it was quite possible to provide some concrete answers to our previous question, which are, synchronization dynamics with coherent phases, collision-induced exchange of velocity, desynchronization dynamics from the anharmonic effect, and periodic oscillations in the strong damping limits. These results are not limited to small amplitude vibrations. Some of these results may be beyond the scope of the Mathieu equation (thus without the Floquet theorem), indicating the necessary investigation of the interacting Mathieu equation in future. These results can be found for other types of inter-particle interactions (including Van der Waals interactions) \cite{Trimby2022Reaching}, in which when $r\rightarrow \infty$, $V(r) = 0$. It is important that some of these predictions can be immediately verified in experiments. 

Finally, this work contribute to the study of the dynamics of trapped ions, which can be generalized to higher dimensions \cite{Porras2006QuantumManipulation, Jordan2019NearGround, Wu2020TwoDimensional, Guo2023SiteResolved}, due to the current research of trapped ions for quantum computations. The interacting Mathieu equation is probably much more important due to the wide range study of this model in dynamical systems in terms of the Floquet theorem in mathematics. It is to be hoped that the experimental progresses for the realization of this model can stimulate the investigation of this interacting Mathieu model from a pure mathematical perspective beyond the quasi-periodic solutions \cite{Geng2007QuasiperiodicSI, Ge2022Quasiperiodic}. Meanwhile, this model is also interesting in the presence of noise at finite temperature with both dissipation and fluctuation, which is connected by the Einstein relation in the Brownian motion between the temperature, damping rate and diffusion constant, which is always presented in experiments when the temperature is not sufficiently low \cite{Wen2023Generalized, Pyka2013Topological,Ulm2013Observation}. 

\textit{Acknowledgments}: We thank Prof. Ping Xing Chen and Prof. Dun Zhou for the valuable discussions about the physical realization of this model and about its possible applications in pure mathematics. This work is supported by Innovation Program for Quantum Science and Technology (No. 2021ZD0301200, No. 2021ZD0303200, and No. 2021ZD0301500) and the Alliance of International Science Organizations (ANSO).

%\appendix
%\input{appendix}

%\newpage
\bibliography{ref}

\end{document}